\documentclass[aip, preprint, groupedaddresses,superscriptaddress]{revtex4-1} 

\usepackage{url}
\usepackage{graphicx}% Include figure files
\usepackage{dcolumn}% Align table columns on decimal point
\usepackage{bm}% bold math
\usepackage{amssymb,amsmath}
\newcommand{\subscript}[1]{\ensuremath{_{\textrm{#1}}}}   % subscript

\begin{document}

\title{Molecular reordering processes on ice (0001) surfaces from long timescale simulations}

\author{Andreas Pedersen}
\affiliation{Faculty of Physical Sciences and Science Institute, University of Iceland, VR-III, 107 Reykjav\'{\i}k, Iceland}
\affiliation{Integrated Systems Laboratory, ETH Zurich, 8092 Zurich, Switzerland}
\email{andped10@gmail.com}
\author{Kjartan T. Wikfeldt}
\affiliation{Science Institute, University of Iceland, VR-III, 107 Reykjav\'{\i}k, Iceland}
\affiliation{NORDITA, AlbaNova University Center, S-10691 Stockholm, Sweden}
\author{Leendertjan Karssemeijer}
\author{Herma Cuppen}
\affiliation{Radboud University Nijmegen, Institute for Molecules and Materials, Heyendaalseweg 135, 6525 AJ Nijmegen, The Netherlands}
\author{Hannes J\'onsson}
\affiliation{Faculty of Physical Sciences and Science Institute, University of Iceland, VR-III, 107 Reykjav\'{\i}k, Iceland}
\affiliation{Dept. of Applied Physics, Aalto University, Espoo, FI-00076, Finland}

%\begin{tocentry}
%\end{tocentry}

\begin{abstract}
We report results of long timescale adaptive kinetic Monte Carlo 
simulations aimed at identifying possible molecular reordering processes on 
both proton-disordered and ordered (Fletcher)
basal plane (0001) surfaces of hexagonal ice. 
The simulations are based on a 
force field for flexible molecules
and
span a time interval of up to
50~$\mu$s at a temperature of 100~K, 
which represents a lower bound to the 
temperature range of Earth's atmosphere. 
Additional calculations using both density functional theory and
an \textit{ab initio} based polarizable potential function
are performed to 
test
and refine the force field predictions. 
Several distinct processes are found to occur readily even at 
this
low temperature, 
including concerted reorientation (flipping) of neighboring surface 
molecules, which changes the pattern of dangling H-atoms, and the
formation of interstitial defects by the downwards motion of upper-bilayer 
molecules. On the proton-disordered surface,
one major surface roughening process
is observed that significantly disrupts the crystalline structure. Despite much longer 
simulation time, 
such roughening processes are not observed on the highly 
ordered Fletcher 
surface which is energetically more stable because of
smaller repulsive interaction 
between neighboring dangling 
H-atoms.
%H-bonds.
However, a more localized process takes place on the Fletcher surface
involving a surface molecule transiently leaving its lattice site.
The flipping process provides a facile pathway of increasing 
proton-order and stabilizing the surface, supporting a predominantly 
Fletcher-like ordering of low-temperature ice surfaces, but
our simulations also show that proton-disordered patches on the surface 
may induce significant local reconstructions. Further, a subset of the 
molecules on the Fletcher 
surface are susceptible to forming interstitial defects 
which might provide active sites for various chemical reactions in the atmosphere.
\end{abstract}

\maketitle

\section{Introduction}
\label{sec:Introduction}

The surface of water ice plays a key role in atmospheric sciences. 
It moderates the temperature on Earth through reflection of sunlight, traps trace gases 
and catalyzes various chemical reactions that are inefficient in the gas phase. 
An important example of the latter is the production of active chlorine species that destroy 
ozone molecules in polar stratospheric regions~\cite{bartels2012ice}.
The molecular-level structure and dynamics at ice surfaces are, however, difficult to
investigate experimentally and a range of fundamental questions remains unanswered. 
At the same time, classical dynamics simulations are challenging because of the 
need for an accurate description of the 
molecular interactions and capturing processes that take place on 
a timescale that is very long compared with vibrational motion.

Ice Ih is the most common ice phase on Earth. In this phase,
the oxygen atoms of 
the water molecules form a hexagonal tetrahedrally ordered lattice. 
The H$_2$O molecules can have a random orientation as long as the ice-rules are 
obeyed~\cite{Bernal:1933kw} -- every H$_2$O donates and receives two hydrogen (H-) bonds 
and there is one H-atom between each nearest neighbor oxygen-oxygen pair.
At moderate cooling below the freezing point, \textit{i.e.}, from 273~K down to about 240~K, the ice surface is 
characterized by a disordered quasi-liquid layer, but the exact temperature at which 
this layer forms and how its thickness depends on temperature 
remains to be determined~\cite{li2007surface}. 
Nonetheless, timescales involved in this temperature range are within reach of 
conventional simulation techniques, and classical dynamics simulations using force fields 
have provided insights into such disordering and pre-melting 
phenomena~\cite{kroes1992surface, Bolton:2000cl,Bishop:2008dh}.
At lower temperature,
where polar stratospheric clouds form, below
$\sim 200$~K, the ice surface is more rigid but 
molecules in the top bilayer are still significantly more mobile than molecules in the 
crystal.
Laser-induced thermal 
desorption measurements on isotopically substituted ice films 
have indicated a mean residence time
in the range of 10$^{-3}$~s to 10$^{-6}$~s at 180-210~K 
before surface molecules diffuse into the 
crystal~\cite{brown1996surface}.
At even lower temperature,
the surface structure and dynamics of ice is relevant for understanding 
chemical processes in polar mesospheric clouds that form at temperature
down to around 140~K.
Ice is also abundant in the interstellar medium where the temperature 
can be as low as $\sim$10~K.
Although the majority of ice under such conditions is believed to be amorphous, observations suggest 
that also crystalline ice is present~\cite{maldoni2003crystalline}. 

\begin{figure}
\centering
\includegraphics[width=.5\textwidth]{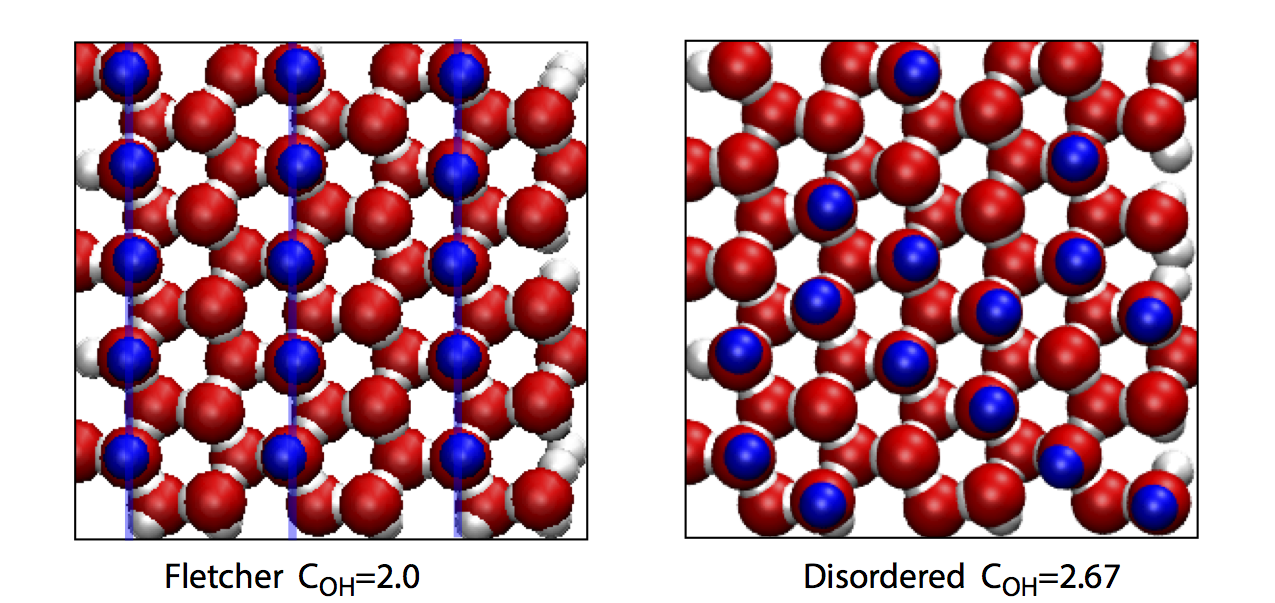}
\caption{(Color online) Ice Ih (0001) surfaces with a (left) Fletcher structure
and a (right) proton-disordered structure.  
Dangling surface H-atoms, 
d-H, are colored blue.  
On the disordered surface there are local regions with 
a high density of nearest neighbor d-H %, marked by shaded blue area, 
while for the Fletcher 
surface all H-atoms are in a local environment with 
only two neighboring d-H. 
}
\label{fig:fig1}
\end{figure}

The 
molecular-level surface structure of ice at low temperature
is still a subject of debate.
Low-energy electron diffraction (LEED)~\cite{materer1997molecular} and 
helium atom scattering experiments~\cite{glebov2000helium} has given strong support for a full-bilayer
termination, \textit{i.e.}, an intact surface bilayer,  of the (0001) surface below 100~K
although the LEED measurements at 90~K indicated sufficiently strong vibrational 
motions of the outermost molecules to render them undetectable~\cite{materer1997molecular}.
If indeed basal plane ice surfaces do not reconstruct the question arises what the ordering is 
of dangling H-atoms (d-H) at the surface, \textit{i.e.}, H-atoms that do not participate in H-bonds. 
The ice crystal is characterized by 
proton-disorder and if the disorder persisted on the surface these d-H would form 
a random pattern. This would lead to a large variety of adsorption sites on the surface, 
with a wide range in binding 
energy and activation energy
for site-to-site
hopping of water admolecules, reminiscent of the surface of an amorphous 
material~\cite{Batista:2001tj}. However, several theoretical studies 
employing both empirical force fields~\cite{Buch:2008ij} and density functional 
theory (DFT)~\cite{Pan:2008ci,Pan:2010ic} to determine minimum energy configurations of
ice surfaces have shown that the ordered Fletcher 
surface~\cite{Fletcher:1992wi} -- where the 
d-H are arranged in a linearly striped pattern, see Fig.~\ref{fig:fig1} -- is the configuration
with lowest surface energy since repulsive nearest neighbor 
interactions between d-H are minimized.
The average number of nearest neighbor d-H,
\begin{equation}
\label{equ:order}
C_{\text{OH}} = \frac{1} {N_{\textrm{OH}}} \sum_{i=1}^{N_{\textrm{OH}}}c_i,
\end{equation}
where $N_{\text{OH}}$ is the total number of d-H and $c_i$ is the number of nearest neighbor d-H surrounding
molecule $i$,
gives a simple and convenient measure of the average repulsive energy between d-H~\cite{Pan:2008ci}.
For the perfect Fletcher 
surface,
$C_{\text{OH}}$=2.0, while $C_{\text{OH}}$
increases with increasing disorder and a completely random arrangement gives $C_{\text{OH}}$$\approx$3~\cite{Pan:2010ic}.
Examples of both surface types are shown in Fig.~\ref{fig:fig1}, 
where $C_{\text{OH}}$ for the disordered phase was determined to be 2.67.  
The Fletcher 
surface,
is, however, entropically unfavorable~\cite{Fletcher:1992wi,Pan:2008ci},
and a mosaic adopting an 
intermediate degree of linear order has been proposed as a lower free energy state~\cite{Buch:2008ij}. 

Even though the detailed surface d-H morphology remains 
undetermined, it is known that the pattern has a large influence on 
the energetics of various types of surface defects~\cite{watkins2011large,Watkins:2010jq} which may play an important role in determining the chemical reactivity of ice surfaces in the atmosphere
as well as in interstellar space. However, obtaining structural and dynamical 
information on surface proton ordering, defect formation, 
and possible reconstructions of ice at low temperature is challenging 
for
conventional simulation techniques due to the long timescales involved
as compared with vibrational periods.
The purpose of the present work is to 
address
this problem by applying the adaptive kinetic Monte Carlo 
(AKMC)~\cite{Henkelman:2001ur} method to ice surfaces.
Recently we demonstrated the ability of AKMC to simulate, 
without biasing the system in any way, the long timescale dynamics of a CO 
admolecule on the basal 
plane of ice~\cite{Karssemeijer:2012ce}.
In the present study,
 the long timescale 
dynamics
of the pristine basal 
plane of ice is simulated using AKMC for both proton-ordered and 
disordered surfaces using the TIP4P/2005f~\cite{Gonzalez:2011eb} force field.
Our AKMC simulations reveal several interesting reordering processes with sufficiently low energy 
barriers to occur readily on a $\mu$s timescale at 100~K.
Interstitial defects can form on both 
disordered and Fletcher 
surfaces when a top half-bilayer molecule 
with a d-H reorients, shifts down and bonds to a second-bilayer molecule. 
On a proton-disordered surface we observe both a collective reorientation of three 
neighboring molecules in the surface bilayer, 
resulting in a stabilization of the surface due to lower repulsion between 
nearest neighbor d-H, and a surface roughening 
process that significantly disrupts the crystalline order in the top 
bilayer, yet still lowers the structural energy. 
Such a major reconstruction was not observed on the Fletcher surface,
despite significantly longer simulation time, although one transient process 
occurred involving a molecule leaving its lattice site and forming a possible precursor 
state to a vacancy defect for around 10~ns before returning 
to its crystalline position.
We then compare the energetics of the reordering processes 
observed using
TIP4P/2005f 
to more advanced approaches, \textit{i.e.}, DFT as well as the \textit{ab initio}-based single-center multipole expansion (SCME) water 
model~\cite{wikfeldt2013scme}, to 
test
and refine the force field predictions.

This paper is organized as follows. In the next section the ice surfaces
and interaction potentials used in this study are described along with 
details of the AKMC method. We then present our results on both proton-disordered and ordered 
Fletcher 
surfaces and compare
the force field predictions to more elaborate interaction models,
and finally give 
concluding remarks in the final section.

%---------------------------------------------------------------------------------------------------------------------
%---------------------------------------------------------------------------------------------------------------------

\section{
%Computational Details
Methodology
}
\label{sec:Simulation}

\subsection{Hexagonal ice samples}

To model and analyze the pristine ice Ih (0001) surface,
two samples 
with different d-H patterns on the surface were constructed.
Both samples contain 360 water molecules arranged into 6 bilayers. 
The construction of each sample started from a 
hexagonal ice crystal
with randomly generated H-bond network according to the method of Buch {\it et al.}~\cite{Buch:1998in}. 
The $c$-axis was chosen along the $z$-direction and periodic boundary conditions were applied in the $x$- and $y$-directions to mimic an infinite surface. 
The ionic and simulation cell degrees of freedom were then optimized using the TIP4P/2005f force field~\cite{Gonzalez:2011eb} 
under the constraint that the ratio between the side-lengths in $x$- and $y$-directions remained constant ($a$ and $c$ are allowed to vary). 
The resulting ratio between $c$ and $a$ is 1.738, which is 6.4\% larger than the ratio for a 
perfect hcp lattice and 6.8\% larger than the experimentally observed ratio for ice Ih~\cite{Rottger:2012im}. 
For the energy-minimized structure,
the lowest two bilayers were then frozen in the 
crystal
configuration and the surface was created by adding a vacuum layer in the $z$-direction. 
This method generated an ice Ih substrate with a disordered d-H pattern. To create a Fletcher 
surface, the same method was applied, but with an initial 
crystal
sample with a 
Fletcher-like 
H-bond pattern between two of the bilayers.
The d-H patterns of both substrates are shown in Fig.~\ref{fig:fig1}. 
The calculated surface energy for the top bilayer is  8.40~meV/\AA$^2$ and 9.13~meV/\AA$^2$ for the Fletcher 
and 
disordered surfaces, respectively, calculated as 
\begin{equation}
\gamma = \frac{1}{A}\left(\sum_{i=1}^{N=60}E_{\textrm{b}}^{(i)} - N \left<E_{\textrm{b}}
%^{(bulk)}
\right> \right) 
\end{equation} 
where the sum runs over the binding 
energy
$E_{\textrm{b}}$ of all molecules in the top 
bilayer, $\left<E_{\textrm{b}}
%^{(bulk)}
\right>$ is the 
cohesive energy of the crystal,
635~meV, 
and $A$ is the surface area, 23.08~\AA\,$\times$ 22.21~\AA.

\subsection{Adaptive kinetic Monte Carlo}

To sample configuration space, the adaptive kinetic Monte Carlo 
method~\cite{Henkelman:2001ur,Pedersen:2010js}
was
used as implemented in the EON software~\cite{chill2014eon}. This involves sampling potential energy minima 
by traversing through saddle points (SPs) on the potential energy surface according to the kinetic Monte Carlo (KMC) algorithm. 
To locate SPs, the iterative minimum-mode following method was applied~\cite{henkelman1999dimer,olsen2004comparison,Pedersen:2011cd}. 
This method uses the lowest eigenvalue mode of the Hessian matrix which was, in this case, estimated by the Lanczos method~\cite{tennyson1986calculation, malek00_7723}. 
The search for a SP starts by slightly displacing the system 
at random
from its initial potential energy minimum, referred to here as reactant state. 
Then, the force component parallel to the minimum-mode is inverted and a climb on the potential energy 
surface up to a SP is conducted by applying an ordinary minimization algorithm.
The two minima separated by the located SP, the reactant and product states, are 
determined by displacing the system along and in the opposite direction of the minimum-mode at the SP followed by 
minimization. Searches for SPs and minima were considered converged when the maximum force acting on any atom
decreases below 1~meV/\AA. 
Transition rates through the located SPs were estimated using Harmonic Transition State Theory (HTST):
\begin{eqnarray}
\label{equ:HTSTRate}
k^{\textrm{HTST}}&=&\nu \exp \left[-\frac{E_{\textrm{SP}}-E_{\textrm{R}}}{k_{\textrm{b}}T} \right]\\
\nu&=&\frac{\prod^D_i\nu_{\textrm{R,}i}}{\prod^{D-1}_i\nu_{\textrm{SP,}i}},
\end{eqnarray}
where ${E_{\textrm{SP}}}$ is the energy of the SP, ${E_{\textrm{R}}}$ is the energy of the reactant configuration, ${\nu_{\textrm{SP,}i}}$ and ${\nu_{\textrm{R,}i}}$ 
are the %non-imaginary 
frequencies of the vibrational modes at the SP configuration excluding the unstable mode and the reactant configuration, respectively,
and $D$ is the number of degrees of freedom.
For a given reactant state, a successful SP search that identifies a unique
SP connecting to a product state is termed a transition and is 
entered into the event table of that state.
When sufficiently many transitions for a reactant state have been found
and the event table is 
considered to be complete enough, the simulation proceeds 
by picking one of the transitions at random with probability proportional to their relative rates. 
Time is then advanced by
\begin{eqnarray}
\label{equ:TimeStep}
\Delta t=-\frac{\ln \mu }{\sum_j k^{\textrm{HTST}}_j},
\end{eqnarray}
where ${\mu}$ is a random number on the interval $(0,1]$ and $j$ runs over all distinct 
transitions in the event table, 
and the system is moved to the product state.

The initial displacements 
to start the SP searches
were made by rotating and translating one 
surface H$_2$O molecule chosen at
random  
in the uppermost bilayer.
The 
displacements were determined by values drawn from
a 
Gaussian distributions having 
a standard deviation of 
0.25 radians for the rotation and 0.25 \AA\ for the displacement.  
The event table for each state
was considered to be complete when a confidence level of 0.99 was obtained as defined by Xu {\it et al.}~\cite{xu08_114104}. 
At this confidence level, an average 
of 438 (521) unique SPs were tabulated for each state visited on the 
disordered (Fletcher) surface, 
while the corresponding total numbers of located SPs 
were on average 1209 (1498).  
The success rate
of the SP searches in finding  
low energy SPs connected to the initial state minimum 
was
57\% (65\%). 
To further enhance the sampling capability of AKMC, an automated coarse-graining
procedure for grouping together states connected by 
fast transitions
was applied~\cite{Pedersen:2012wq,jonsson2011simulation}. 
The temperature was set to 100~K in the AKMC simulation, but the energy barrier and pre-exponential factor 
for each transition were saved and 
could
in principle be used to simulate the system at other temperature
within the HTST approximation.

\subsection{Interaction potentials}
In all 
the
AKMC simulations,
the inter- and intramolecular interactions were modeled with the TIP4P/2005f potential~\cite{Gonzalez:2011eb}, 
which  is a flexible version of the TIP4P/2005 potential~\cite{Abascal:2005ka}.  
The system is subject to periodic boundary conditions and all interactions are smoothly 
truncated at distances between 9 and 10~\AA, 
based on the separation between the centers of mass of the molecules. 

Empirical potentials such as TIP4P/2005f have been parametrized to fit various 
experimental data on 
extended phases such as liquids and/or crystals
and are, therefore, not 
necessarily transferable to different environments such as an ice surface. 
To compare with previous DFT calculations on ice surfaces the vacancy 
formation energies for each of the 30 molecules in the upper half of 
the top bilayer for both the disordered and Fletcher surfaces were computed.
%We therefore compare with previous DFT calculations on ice surfaces by computing vacancy 
%formation energies for each of the 30 molecules in the upper half of 
%the top bilayer for both the disordered and Fletcher surfaces. 
One by one each surface molecule was removed and the vacancy defect surface relaxed.
The vacancy formation energy is defined as the energy difference between 
the final relaxed defect surface and the initial defect-free surface.
A broad distribution 
of vacancy formation energies is found, ranging from 299 to 421~meV and  
from 179 to 422~meV for the Fletcher and disordered surfaces, 
respectively. The corresponding average energies are 365 
and 311 meV, reflecting the larger stabilization of molecules on the 
Fletcher surface and thus larger vacancy formation energies. 
Both the averages and the magnitude of the variation are in rather good agreement
with previous DFT calculations~\cite{watkins2011large}, 
where the PBE functional gave vacancy formation energies ranging from 
around 200~meV to 550~meV, and two dispersion-corrected functionals 
as well as the hybrid PBE0 functional gave consistent results 
(see Supplementary Material of ref.~\cite{watkins2011large}).
This suggests that TIP4P/2005f provides a reasonable description of the surface energetics. 
However, DFT predicted the 
existence of even higher energies than found here by TIP4P/2005f. This discrepancy can 
be easily understood since higher vacancy formation energies were found to correlate strongly 
with enhanced molecular dipole moments, but dipole polarizability is neglected in 
classical force fields such as TIP4P/2005f.

To further reinforce our conclusions presented below 
based on AKMC simulations employing 
TIP4P/2005f we performed additional calculations using DFT and a recent 
\textit{ab initio}-based polarizable water potential, the single-center multipole expansion model 
(SCME)~\cite{wikfeldt2013scme}, on structures 
generated by the AKMC simulations. 
SCME describes water molecules as interacting static electric multipoles, with 
the expansion of the electric field going up to the hexadecapole, and 
many-body effects are treated by polarizable dipole and quadrupole moments.
Apart from the rigorous description of the electrostatic interactions, 
SCME also includes dispersion coefficients from quantum chemistry calculations 
and a density-dependent short-range repulsion interaction.
For DFT calculations a
van der Waals density functional (optPBE-vdW~\cite{klimevs2010chemical}) was used to 
include non-local dispersion 
interactions which are important for a realistic description of ice~\cite{santra2013accuracy}. 
This functional was shown to accurately describe the energetics of water clusters~\cite{klimevs2010chemical},
and the density of ice Ih is predicted to within 2\% accuracy compared to experiments 
while it overestimates the cohesive energy by 58~meV/H$_2$O, 
\textit{i.e.}, around 10\%~\cite{santra2013accuracy}.
SCME also predicts 
the equilibrium density
of ice to 
be 
within 2\% of experiments and, furthermore, 
accurately reproduces the experimental cohesive energy~\cite{wikfeldt2013scme}.

When evaluating DFT and SCME 
energies of
the stable structures resulting from the AKMC simulations, the size of the cell 
was first rescaled, since neither optPBE-vdW nor SCME predict the same lattice parameters as TIP4P/2005f, 
and all ionic positions were then relaxed.
The lattice parameters of SCME were reported in ref.~\onlinecite{wikfeldt2013scme} and those of optPBE-vdW in ref.~\onlinecite{santra2013accuracy}. 
The DFT calculations were performed with the CP2K/Quickstep 
software~\cite{vandevondele2005quickstep}, where 
the electronic structure is described with a dual Gaussian and plane-wave 
basis set. Core electrons were described by the Goedecker, Teter and 
Hutter (GTH) pseudopotentials~\cite{goedecker1996separable}, valence electrons were expanded in a 
molecularly-optimized polarized triple-$\zeta$ basis 
(TZVP-MOLOPT-GTH) and the cutoff energy of the auxiliary plane wave basis 
set was set to 340~Ry. 

\begin{figure}
\centering
\includegraphics[width=.5\textwidth]{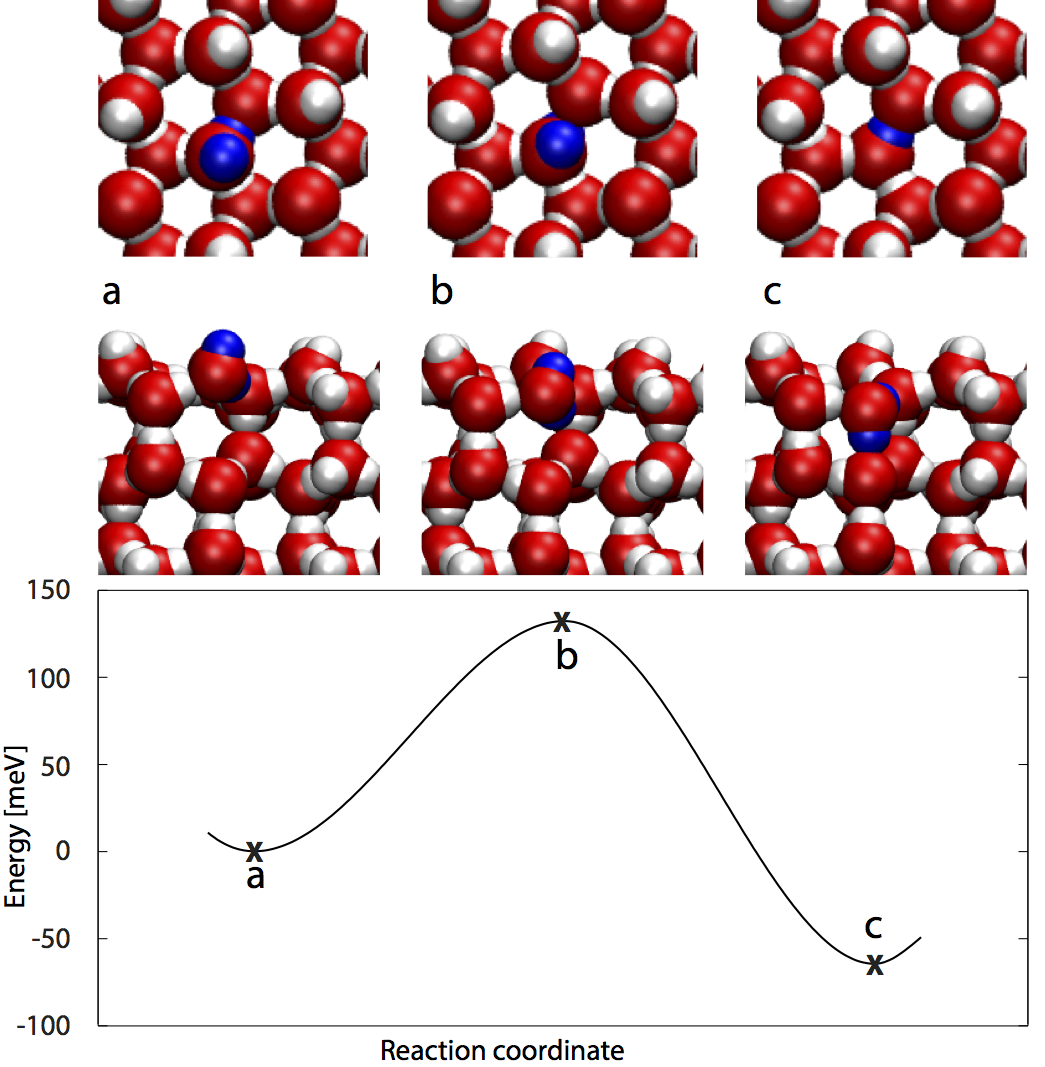}
\caption{(Color online) Formation of an interstitial defect on the proton-disordered surface 
  where a top-half surface bilayer molecule shifts down into the surface.
  Upper row of insets are top-views and lower row are side-views.
  The energy of the pristine disordered surface defines the zero point on the energy scale. 
  The H-atoms of the moving water molecule are colored blue. 
  In this as well as following figures crosses mark computed 
  energy of the stable and 
  saddle point states found by AKMC, the continuous line is a guide to the eye and the insets only show the topmost bi-layer.}
\label{fig:fig2}
\end{figure}
%---------------------------------------------------------------------------------------------------------------------

\begin{figure}
\centering
\includegraphics[width=.5\textwidth]{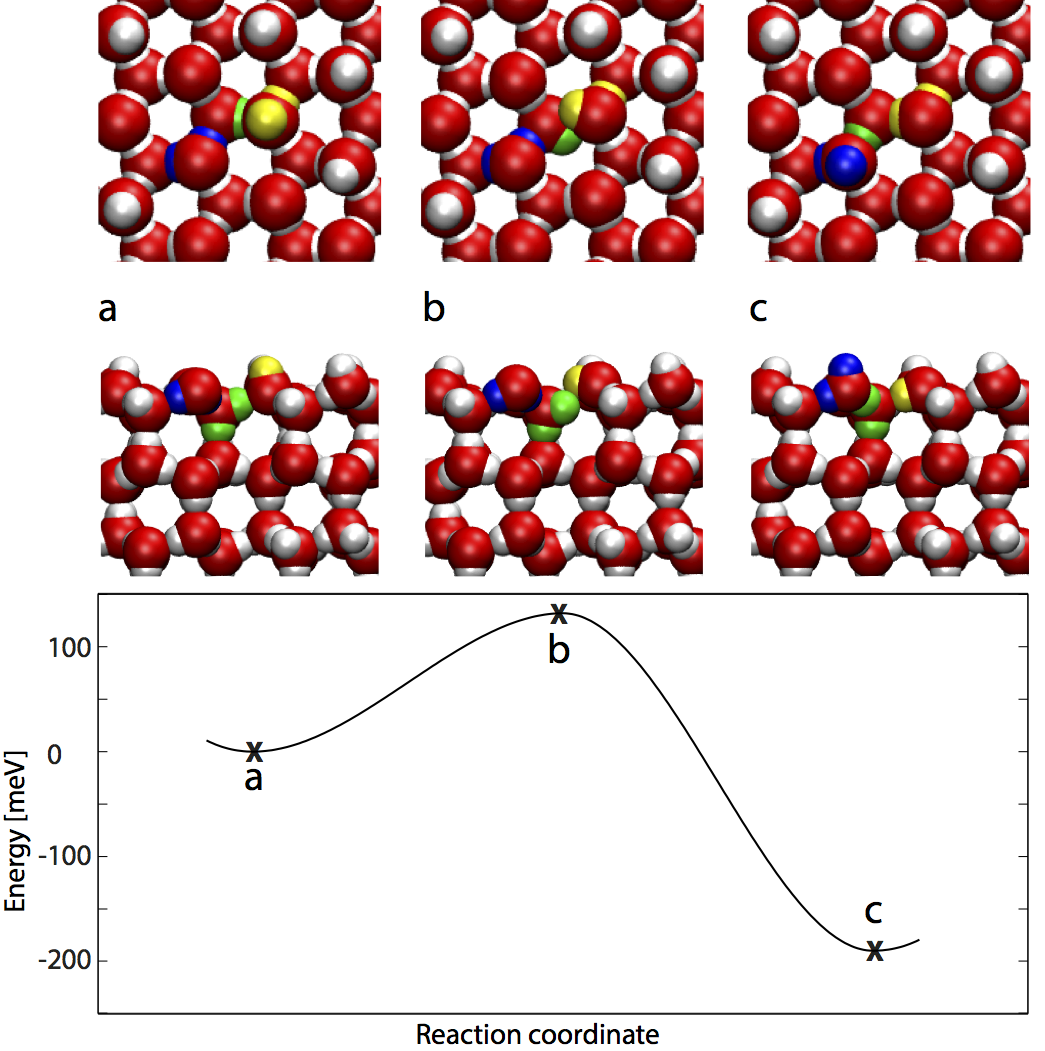}
\caption{(Color online) Concerted flipping of three molecules resulting in a reduced
clustering of dangling H-atoms on the proton-disordered surface.
Upper row of insets are top-views and lower row are side-views.
Molecules participating in this reordering have differently colored 
H-atoms, and the energy of the pristine disordered surface defines the zero point on the energy scale.  
}

\label{fig:fig3}
\end{figure}
%---------------------------------------------------------------------------------------------------------------------

\section{Results}
\label{sec:Results}
We now discuss the results of the AKMC simulations for the 
two models of the (0001) surface.
Starting from both a disordered and a Fletcher 
surface the 
long timescale dynamics at 100~K was simulated.  
For the disordered surface,
the simulation was terminated after a surface roughening 
process occurred at 0.11~$\mu$s when a total of 15330 unique transitions had been 
determined.
The simulation of the Fletcher 
surface was continued up to 49.4~$\mu$s, at which time
a total of 112536 unique transitions had been found.

\subsection{Proton-disordered surface}
\label{sec:disord}
Three types of reordering processes were observed on the disordered surface.  
Two of these processes have a strongly localized characater and 
involve the crossing of a single SP,
%and are thus represented
%by a single transition in the event table,
while the third is a series of transitions resulting in a reordering of several 
surface water molecules and a roughening of the surface.   

\subsubsection{Interstitial formation}
The simplest type of process
observed was
the creation of an interstitial defect. 
Although the event tables constructed 
from the SP searches
contained several such transitions, 
only one was chosen
as part of the time evolution of the system over the simulated time period of 
0.11~$\mu$s.
This transition is depicted in Fig.~\ref{fig:fig2} where the H-atoms of the affected 
molecule are colored blue. A surface water molecule with a d-H rotates 
in its molecular plane and shifts downwards by 1.55~\AA\ 
to reach a configuration where it becomes an interstitial defect within the 
surface bilayer, forming
a H-bond to a molecule in the top part of the second bilayer.  
The distance between the oxygen atoms in the newly formed H-bond is 2.83~\AA, 
which is longer than the 
equilibrium distance of 2.73~\AA\ in the crystal.
In the final structure the number of d-H is decreased by one and the number 
of nearest neighbor d-H interactions decreases by four since the molecule was initially in a region 
with a cluster of d-H, see Fig.~\ref{fig:fig2}.  
For the transition to take place a barrier of 130~meV must be surpassed and the resulting change in energy is a lowering of 60~meV.   
The pre-exponential factor for this process was calculated to be $1.3 \cdot 10^{13}$~s$^{-1}$. 

It might be expected that the molecule undergoing the transition to become an interstitial is 
in an energetically unfavorable geometry, particularly since it has four 
nearest neighbor d-H. % (see eq.~\ref{equ:order}). 
However, while the average binding energy, $E_\textrm{b}$, of molecules in the upper part of the top bilayer is 490~meV before the interstitial formation, 
$E_{\textrm{b}}$ of the affected molecule is 493~meV, close to the average. 
After becoming an interstitial, the binding energy of the molecule is significantly increased to 584~meV.
On the other hand, the binding 
energy
of the nearest neighbor molecules on the surface 
is
only 
slightly increased, 
by
at most by 8~meV due to the reduction in repulsive d-H interactions. 
A small destabilization of molecules surrounding the newly formed interstitial brings the total energy lowering 
down to 60~meV. The stabilization of the surface 
thus appears to be
a consequence of the 
increased number of H-bonds formed when the initially d-H molecule rotates 
and shifts downwards to donate an H-bond to a second-bilayer molecule.

Calculated energy differences between reactant and product states of
the interstitial formation process 
by DFT/optPBE-vdW and the polarizable SCME potential differ somewhat from the TIP4P/2005f value, 
as can be seen in Table~\ref{tab:dft}. SCME predicts qualitatively similar energetics with 
an energy lowering of 25~meV, while DFT predicts an energy increase of 64~meV. Since applying 
higher levels of theory for a system of this size would be prohibitively expensive, 
it can only be concluded at this point that the interstitial configuration is a local energy 
minimum with an energy close to the pristine surface.
A more extensive discussion of the comparison between interaction potentials will be presented in section~\ref{sec:intpots} below.

\subsubsection{Concerted reorientation}
\label{sec:flip}
Another possible process on the disordered surface revealed by the AKMC simulation
is a concerted reorientation of three molecules which shifts a d-H from one surface 
molecule to another. This process is analogous to the ``relay'' mechanism observed by Bishop 
{\it et al.}~\cite{Bishop:2008dh} in classical dynamics simulations using
the NvdE six-site potential at 230~K, where it occurred 
on a ns timescale. Here, we find that it can occur also at 100~K on a $\mu$s timescale.
Similarly as for the interstitial formation, several instances of this type of flipping 
process were found in the event tables constructed 
from the SP searches but only one was chosen 
as part of the simulated time evolution of the system.
This transition is depicted in Fig.~\ref{fig:fig3}, where the affected molecules have H-atoms 
colored yellow, green and blue, respectively. 
In this reordering process a d-H of an upper-half bilayer surface molecule (yellow) in a 
dense d-H region rotates such that the d-H forms a new H-bond to a lower-half 
bilayer molecule (green). In the SP configuration the upper-half bilayer molecule has rotated
 by 120$^{\circ}$ while the second molecule has only partially rotated by around 60$^{\circ}$. 
Going from the SP configuration to the final state, 
the lower-half bilayer molecule rotates another 60$^{\circ}$ to donate an H-bond to a second upper-half bilayer molecule (blue), which in turn
rotates and becomes a molecule with a d-H. It can be seen that the number of nearest neighbor
d-H interactions on the surface is reduced by two by swapping a d-H
from a clustered d-H environment to a region with fewer d-H. Through this rearrangement 
the surface energy decreases by 190~meV by surpassing  an energy barrier of 130~meV.  
The pre-exponential factor for this process is determined to be $1.1 \cdot 10^{14}$~s$^{-1}$. 

The binding energy of all three molecules involved in the flipping process increases.
The first molecule, initially with a d-H, is in an unstable geometry with 
$E_{\textrm{b}}$=438~meV, 52~meV below the average surface $E_{\textrm{b}}$. After the transition
it is 481~meV, close to the average of 490~meV. The second molecule in the lower-half bilayer is stabilized by 30~meV, and the binding energy of 
the third molecule that rotates to become a d-H molecule increases
from $E_{\textrm{b}}$=452~meV to $E_{\textrm{b}}$=483~meV, again moving closer to the average $E_{\textrm{b}}$.
Since in the final configuration the number of d-H remains unchanged,
the energy lowering is primarily due to the smaller number of repulsive 
nearest neighbor d-H interactions.

Table~\ref{tab:dft} shows that DFT/optPBE-vdW and SCME predict similar energetics for 
this
process. The predicted stabilization is qualitatively
consistent with TIP4P/2005f but more exothermic by around 100~meV for SCME and by around 40~meV for DFT. 

\begin{figure}
\centering
\includegraphics[width=.5\textwidth]{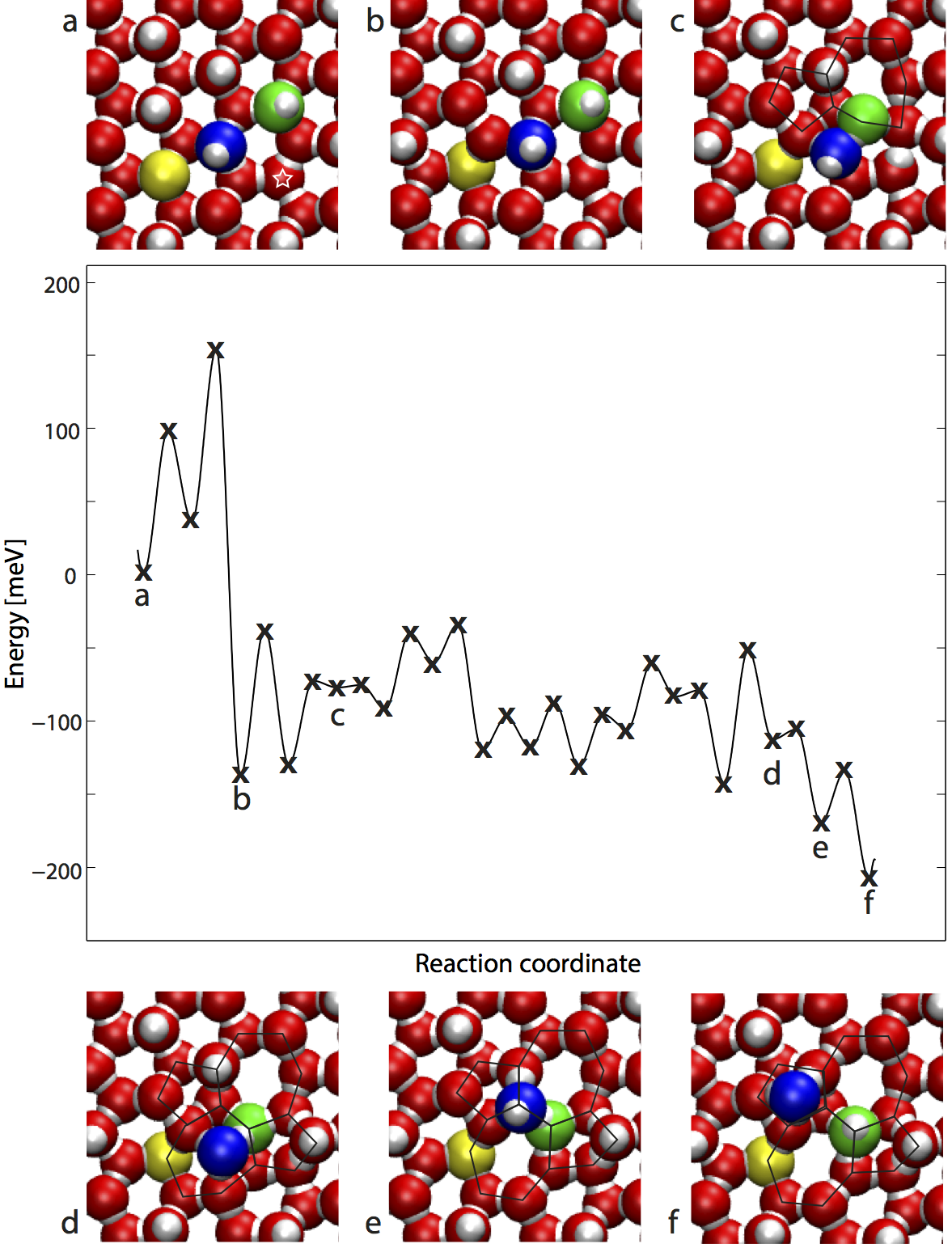}
\caption{(Color online) Roughening process on the proton-disordered surface. 
  Insets are top-views.
  As the surface roughens it looses its perfect planar structure, a vacancy is formed in the surface bi-layer 
  and an admolecule emerges on the surface. Oxygen atoms of the most active 
molecules are colored differently, and
  the energy of the initial disordered surface with one interstitial defect, marked by white star in panel a, 
  defines the zero point of the energy scale.}
\label{fig:fig4}
\end{figure}

\subsubsection{Surface roughening}
The third structural change causing an energy lowering results 
from a series of 15 
transitions
starting from a surface with an interstitial 
defect, shown in Fig.~\ref{fig:fig4},
where the interstitial is marked by a white star. 
A distinct difference as compared 
with the previously discussed transitions is that this process
causes a 
disruption of the hexagonal ordering of the
O-atoms, leading to a large increase in structural entropy.  
The 
activation energy
-- the energy of the highest SP along this 
path
-- is 150~meV and the final reduction in energy amounts to 210~meV. 
The calculated 
values of the
pre-exponential factor
span a range from $5.4 \cdot 10^{12}$~s$^{-1}$ 
to $2.3 \cdot 10^{14}$~s$^{-1}$.
In the first 
transition, a molecule from the upper-half
surface bilayer, yellow in Fig.~\ref{fig:fig4}, 
shifts down by 1.2~\AA\ while its H-bonded partner from the lower-half 
surface bilayer shifts up by 1.2~\AA.
We will refer to this as
a ``ripple'' transition. 
Neither molecule has a d-H, and the only change in 
H-bond topology that takes place is that a third molecule from the second 
bilayer swaps its donating H-bond from the upward-moving to the downward-moving molecule.
This destabilizes the surface slightly by 30~meV. 
Subsequently, a flipping transition occurs,
analogous to the one discussed above, which reduces the number of 
nearest neighbor d-H interactions by 
two and results in a large drop in energy by around 172~meV and both transitions are visualized
in Fig.~\ref{fig:fig4}, a$\rightarrow$b. 
During the following two transitions, b$\rightarrow$c, several concerted changes in 
H-bond connectivity take place between the initial interstitial defect  
and the position where the ripple transition took place.
As a consequence of this reordering, the H-bonds within a hexagonal ring are stretched and two
others rings change from being hexagonal to being pentagonal and heptagonal. 
This  
is
similar to the Stone-Wales defect~\cite{stone1986theoretical} observed 
in graphene, another hexagonal system. In graphene, and also in carbon nanotubes, 
the formation of these kind of defects are known to play a key role during the first steps 
of the melting transition~\cite{zakharchenko2011melting}. Although such carbon-based
systems 
are very different from hexagonal ice in many respects, it is interesting that 
similar defects are formed here in the AKMC simulation
and it may indicate that they play a role in the premelting 
of hexagonal ice.
In the following nine transitions, c$\rightarrow$d, one molecule (blue) moves upwards and becomes an admolecule.  
The vacancy is accommodated in the surface by the formation of
two additional five and seven membered rings around the green molecule in Fig.~\ref{fig:fig4}.  
The final two transitions, d$\rightarrow$e$\rightarrow$f, reveal a possible migration mechanism for the admolecule that formed.  
To reach the final state, f, the admolecule rotates and shifts to break 
the two donating H-bonds in state d and form two new H-bonds in state f. 
In the final state the number of d-H has decreased by one and the number of 
nearest neighbor d-H interactions by four compared to the initial state, a. 

This process, which is composed of 15 distinct 
transitions,
disrupts the crystalline order of the surface to a surprisingly 
large extent and 
it is important to check its feasibility when the molecular interactions are described using the alternative methods,
DFT/optPBE-vdW and SCME. 
As seen in Table~\ref{tab:dft}, SCME predicts energetics similar to
TIP4P/2005f with an energy lowering of 177~meV, while DFT also
predicts a stabilization but of smaller magnitude, around 62~meV. This 
roughening of the hexagonal lattice
thus appears to be a realistic 
process that might take place
on proton-disordered ice surfaces at low temperature. 

% -------------------

\begin{figure}
\centering
\includegraphics[width=.5\textwidth]{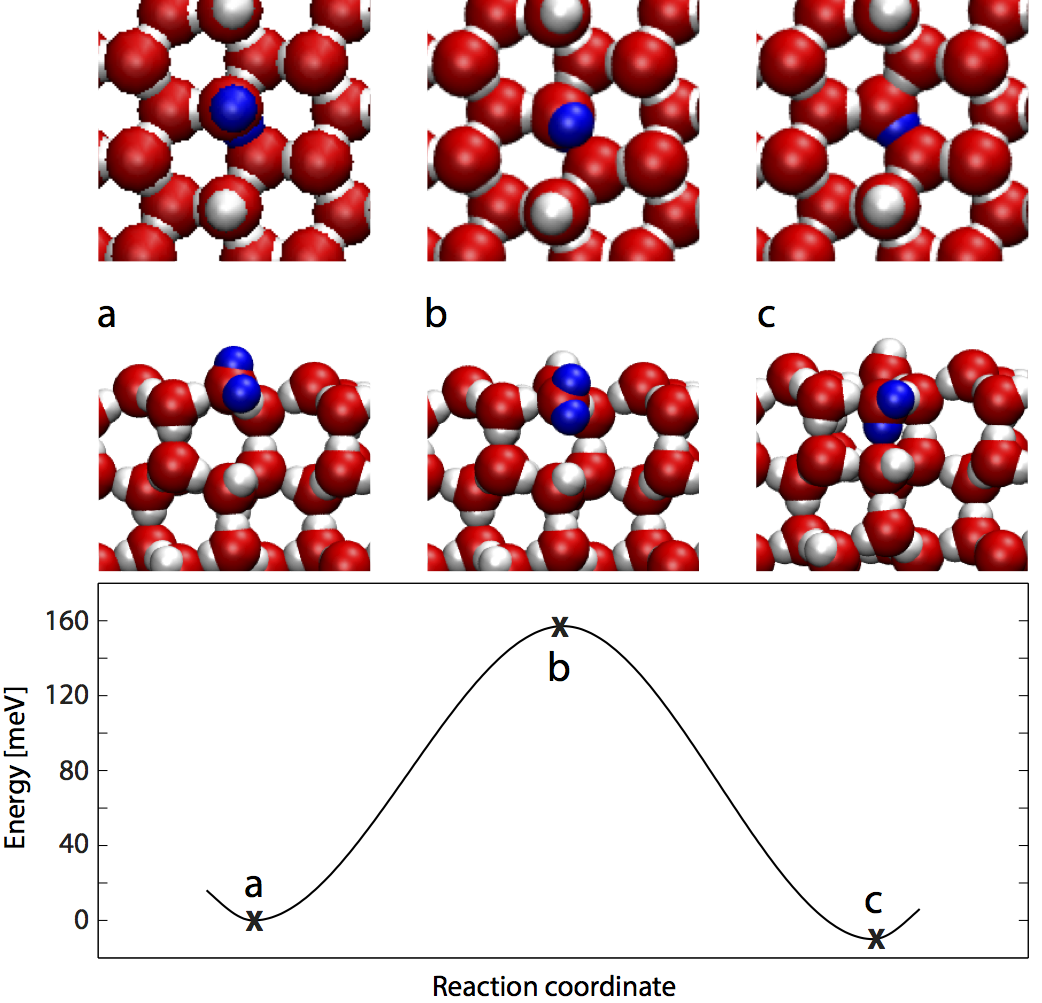}
\caption{(Color online) Formation of an interstitial defect on the perfect Fletcher
  surface
  where a top-half surface bilayer molecule shifts down into the surface.
  Upper row of insets are top-views and lower row are side-views.
  The energy of the pristine surface defines the zero point for the energy scale. 
  The H-atoms of the moving water molecule are colored blue.}
\label{fig:fig5}
\end{figure}

\subsection{Fletcher surface}
The AKMC simulation starting from a perfectly ordered Fletcher
surface
spanned a total of 49 $\mu$s 
and revealed two main types of processes.
As in the case of the proton-disordered surface, interstitial defects can be created 
in single transitions, involving just one SP.
Around 10\% of the surface molecules shift frequently  
between the normal and interstitial positions. 
A
process where one molecule temporarily leaves its lattice site, forming a possible precursor 
state to a vacancy defect on the surface,
was also observed.

\subsubsection{Interstitial formation}
One representative example of an interstitial formation transition is shown in Fig.~\ref{fig:fig5}.  
For the process to take place, the system must overcome an energy barrier of 160~meV and 
the corresponding pre-exponential factor was calculated to be $3.0 \cdot 10^{13}$~s$^{-1}$.  
The barrier for interstitial formation is thus 30~meV larger than for the disordered surface, 
the pre-exponential factor is smaller by a factor of 3.7 and the stabilization of the surface is only 10~meV.
By breaking up one of the stripes of d-H and 
directing a
d-H into the surface,
the system decreases the number of d-H by one and the nearest neighbor d-H interactions by two.  
The mechanism for the creation of the interstitial defect is the same as for the 
disordered surface where a water molecule rotates in the molecular plane. In this transition the molecule 
shifts down by 1.38~\AA\ which is slightly shorter than for the corresponding 
transition on the disordered surface.
Similarly to the disordered surface, the binding energy of the displaced molecule increases from 489~meV, close 
to the average binding energy of upper half top bilayer molecules of 496~meV, to 555~meV.
Other interstitial formation transitions with similar energetics take place 
during the simulation.
Out of the
30 molecules in the upper half surface bilayer, three are particularly prone to becoming interstitial defects and 
they frequently switch between the normal and defect positions, spending a sizable fraction of the 
simulated time
as defects. The average binding energy of these three molecules is 493~meV, close to 
the average $E_{\textrm{b}}$. 
Other factors than the binding energy, such as details of the ordering of H-bonds between 
the first and second bilayers, thus appear to determine the propensity of surface molecules to become interstitial defects.

Considering the previously assumed stability of the Fletcher
surface~\cite{Buch:2008ij,Pan:2008ci,Pan:2010ic},
the slightly lower 
energy of the resulting interstitial defect surface compared to the pristine surface is interesting. 
We find that SCME predicts a very small stabilization, -5~meV, similar to TIP4P/2005f, while DFT/optPBE-vdW 
predicts a destabilization 
of
145~meV. Taking into account also the 
interstitial formation on the disordered surface, there thus appears to be a consistent trend where 
the TIP4P/2005f force field favors the formation of interstitial defects in order to reduce the number of d-H, 
while DFT disfavors it and the \textit{ab initio}-based SCME potential 
gives results that are in between the other two.

\subsubsection{Vacancy formation}
\begin{figure}
\centering
\includegraphics[width=.5\textwidth]{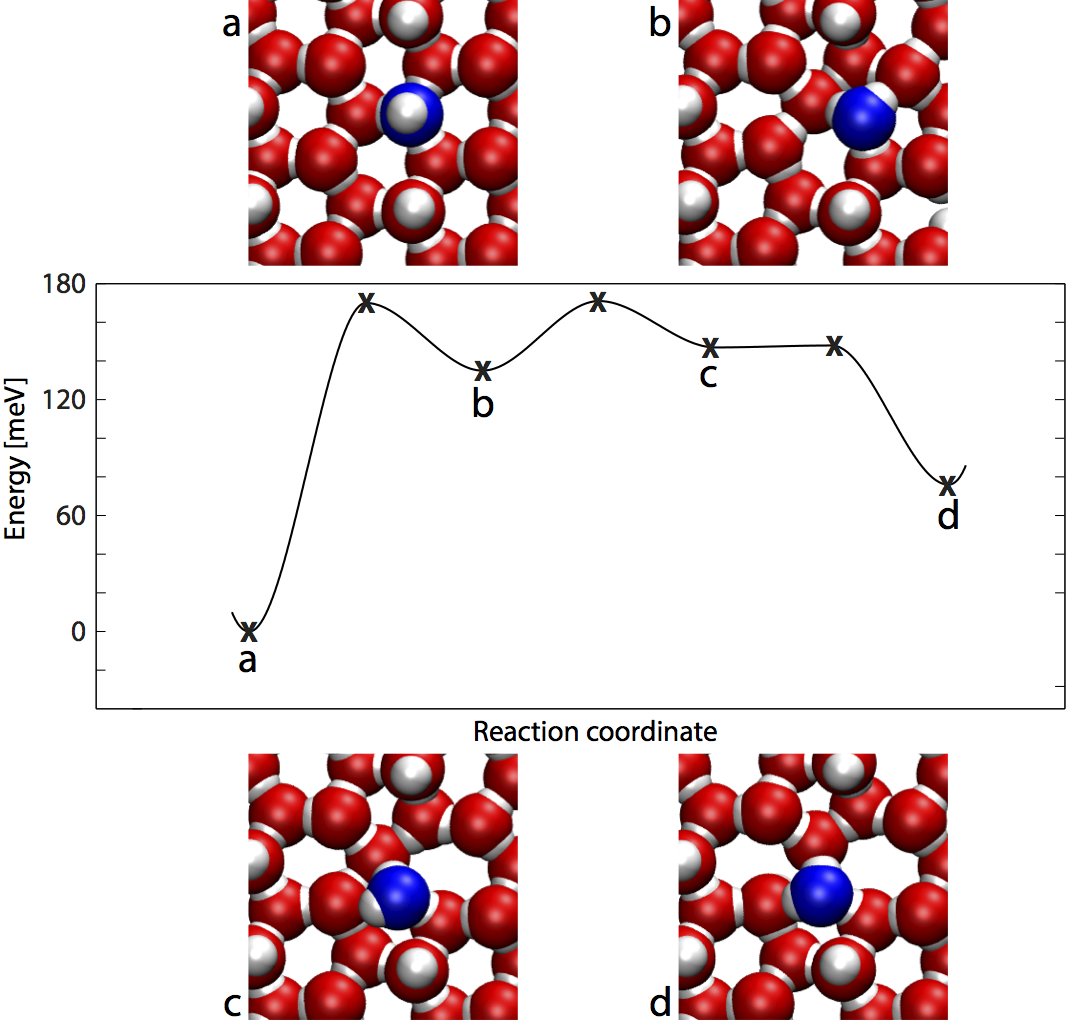}
\caption{(Color online) Possible first series of transitions in the formation of a vacancy defect 
  on the Fletcher
  surface. 
  Insets are top-views.
  The oxygen atom of the active molecule is colored blue. 
  This process leads to a structure where one molecule has left its crystalline site, 
  forming two heptamer and one pentamer rings,
  but in the AKMC simulation the molecule returned to its crystalline site after $\sim$10~ns.}
\label{fig:fig6}
\end{figure}

The second type of process on the Fletcher surface can be viewed as the first 
series of transitions towards the formation of a surface vacancy, 
shown in Fig.~\ref{fig:fig6}. This process starts with the rate-limiting transition
where a molecule (blue) rotates its d-H into the surface plane forming a H-bond to
a neighboring upper-bilayer molecule with
a dangling oxygen lone-pair. The barrier for this transition is 170~meV,
the pre-exponential factor is $3.6\cdot 10^{13}$~s$^{-1}$, and the product state, b, is 
135~meV less stable than the initial state. Through the following two transitions,
b$\rightarrow$c$\rightarrow$d,
the molecule leaves its lattice site and finds a more stable position 
which is, however, 75~meV less stable than the initial state, a. An analysis of the 
binding energy of the top bilayer molecules reveals that the major contribution 
to the destabilization is the less favorable position of the moving molecule which 
leaves a strongly bound site, $E_{\textrm{b}}$=503~meV, to enter a less favorable site, $E_{\textrm{b}}$=451~meV,
 in the final state. No further reordering transitions took place in the 
final state. Instead, the affected molecule returned to its initial crystalline position 
after around 10~ns.

For this process,
the SCME and DFT/optPBE-vdW calculations 
predict a destabilization by 191 and 334~meV, respectively. Again, the 
more disordered nature of the final state leads to a larger energetic preference 
in DFT for the 
perfectly ordered Fletcher
surface
as
compared to the TIP4P/2005f force field, 
with SCME 
giving results that are in between the other two.

%\onecolumn
\begin{table}
\caption{Energy difference, $\Delta E \equiv E_{\textrm{product}} - E_{\textrm{reactant}}$, (in meV) between initial and final states of processes found by AKMC simulations using 
TIP4P/2005f, DFT/optPBE-vdW and SCME. 
}
\begin{tabular}{l | c c c c}
Process/Surface & TIP4P/2005f & DFT & SCME \\
\hline
Interstitial/disordered    & -60 & 64  & -25    \\
Flip/disordered            & -190 & -230 & -287  \\
Roughening/disordered      & -210 & -62 & -177  \\ \hline
Interstitial/Fletcher      & -10 & 145  & -5 \\
Vacancy/Fletcher           & 75  & 334    & 191 
\label{tab:dft}
\end{tabular}
\end{table}
%\twocolumn

%---------------------------------------------------------------------------------------------------------------------
%

\subsection{Comparison of disordered and Fletcher surfaces}
\label{sec:histo}
\begin{figure}
\centering
\includegraphics[width=.5\textwidth]{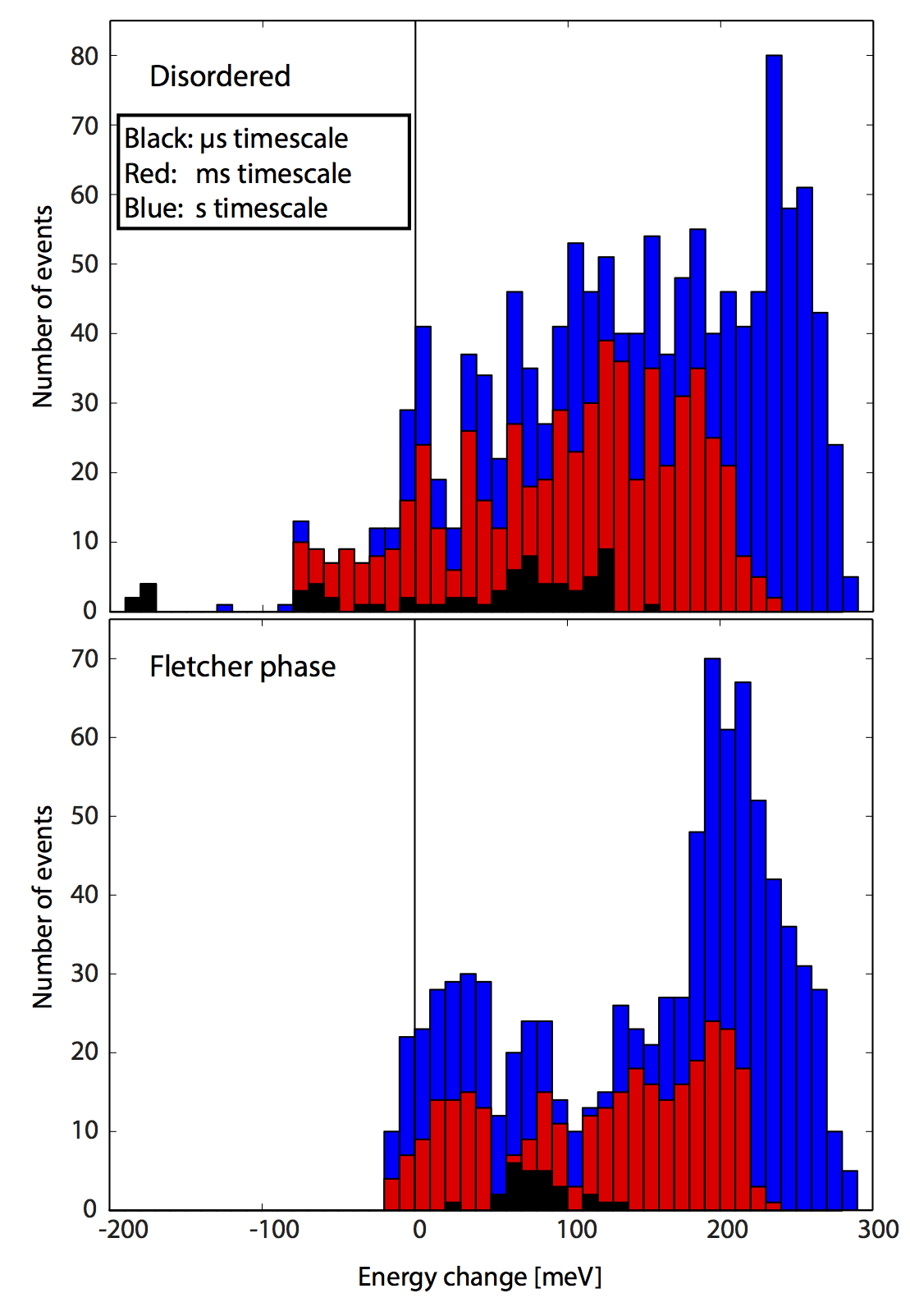}
\caption{(Color online) Histograms of energy changes for transitions found by AKMC extracted from 
all transition tables for the 
proton-disordered and Fletcher
surfaces. Three classes of transitions are shown corresponding to $\mu$s, ms and 
s timescales, respectively. The only transitions included are those 
that can take place from pristine surfaces with 15 d-H or a surface characterized by a single interstitial defect.
 }
\label{fig:fig7}
\end{figure}

The reordering processes discussed above are those 
selected
by the 
AKMC
algorithm 
from all the transitions identified
by the SP searches for the proton-disordered and Fletcher
surfaces.
These 
transitions
only constitute a small fraction of all 
the transitions entered 
in the event tables and additional 
information can be extracted by inspecting also
the
states which were not visited in the 
time evolution generated by the AKMC simulation. 
Figure~\ref{fig:fig7} shows histograms of all 
transitions
in the event tables that are likely to occur within 
specific timescales ordered by their relative energy change $\Delta E \equiv E_{\textrm{product}} - E_{\textrm{reactant}}$. 
To obtain a clear comparison between the disordered and Fletcher
surfaces,
we only considered processes with reactant 
states which are either perfectly crystalline or contain just a single interstitial defect. 
This gives 8 (5) possible initial states and 4056 (2635) 
transitions
in total for the disordered (Fletcher) surface. 
The three classes of 
transitions
shown in Fig.~\ref{fig:fig7} correspond to those that can take place 
on 
a timescale $\tau$ %that is smaller than 
of $\mu$s, ms or s.
The timescale is
obtained from the 
computed HTST rate
as $\tau=1/k^{\textrm{HTST}}$.
These three classes contain 69 (26), 600 (315) and 1288 (877)
transitions each for the disordered (Fletcher) surface.

The disordered surface contains a notably 
larger number of possible transitions that lower the energy 
than
the Fletcher
surface. Even on 
a
$\sim$$\mu$s
timescale,
a range of processes that lower the energy
can occur. 
The transitions with
$\Delta E$$\approx$-180~meV
correspond to flipping transitions 
and are similar to the one
discussed in section~\ref{sec:disord}
and occur on several of the 8 unique initial states of the disordered surface chosen for this analysis. 
The roughening process shown in 
Fig.~\ref{fig:fig6} is not included here since it 
involves a series of transitions.
In contrast, $\sim$$\mu$s processes on the Fletcher 
surface
all increase the energy. 
A closer inspection of the processes marked by black bars in Fig.~\ref{fig:fig7}, lower panel, 
reveals that they involve small perturbations to the surface H-bonds with 
small
molecular displacements. 
The processes accessible on 
a
longer timescale
on the Fletcher surface 
and
involve small energy changes are dominated by interstitial formation transitions as well as ripple transitions.

This comparison of possible transitions found by the SP searches in AKMC, 
shown in Fig.~\ref{fig:fig7}, demonstrates that the surface order present on the Fletcher
surface
greatly reduces the number of molecular reordering processes that lower the
energy. Thus, even though the difference in surface energy between the disordered and Fletcher
surfaces is only around 0.73~meV/\AA$^2$, or about 9\%, the linear ordering of d-H on the Fletcher surface 
has a large effect on the energetics of possible reordering processes.

\subsection{Comparison of interaction potentials}
\label{sec:intpots}

The AKMC simulations using the TIP4P/2005f 
force field 
identified several processes that can take place on the basal plane  
surface of ice Ih 
at low temperature, and the 
viability of these has been tested
using more elaborate approaches, \textit{i.e.}, DFT/optPBE-vdW and the SCME 
\textit{ab initio}-based model.
Some conclusions can be drawn from this comparison.
Firstly, it is worth highlighting that the product states found by AKMC with the TIP4P/2005f force field
are
also 
stable local energy minima when DFT/optPBE-vdW and the SCME potential are used to compute interatomic forces. 
After rescaling the simulation cell to conform to the DFT/optPBE-vdW or SCME equilibrium lattice constants and relaxing the 
atomic coordinates, it is found that the geometrical changes occurring in the 
transitions
found by AKMC are 
similar 
for the three different descriptions of the molecular interactions.
For instance, in the two interstitial formation processes discussed above, 
on the disordered and Fletcher
surfaces, TIP4P/2005f predicts that the affected molecule shifts downwards by 1.55~\AA\ 
and 1.38~\AA, respectively. The corresponding numbers from DFT/optPBE-vdW are 1.49~\AA\ 
and 1.33~\AA\ while SCME predicts 1.60~\AA\ and 1.38~\AA.
This adds confidence in the TIP4P/2005f force field as a reliable model for the description of 
the interaction between water molecules at
ice surfaces and suggests that
the reordering processes discovered in 
the
AKMC simulations are realistic.

Some deviations are nonetheless seen in the energetics.
In the case of the concerted flipping process on the disordered surface, there is good qualitative agreement between TIP4P/2005f, 
DFT/optPBE-vdW and SCME, all predicting a stabilization of the surface by 190-290~meV, see Table~\ref{tab:dft}. The 
energetics of this process is dominated by the electrostatic repulsion between dangling 
H-atoms
on the surface. 
Qualitative disagreement arises in the case of the interstitial formation process on both the disordered and the Fletcher 
phase surfaces,
where TIP4P/2005f and SCME predict a stabilization while DFT/optPBE-vdW gives a higher energy for the less ordered 
final state. 
In these final states the local environment around the interstitial molecule is of higher density than in the
ice crystal; the H-bonds are more bent compared to the crystal and dispersion (van der Waals) interactions 
become stronger. The energetics are dictated by a balance between electrostatic and 
dispersion interactions and the interstitial geometry thus presents a challenge to interaction potentials for water.
\textit{A priori} it cannot be assumed that DFT/optPBE-vdW is significantly more accurate than TIP4P/2005f or SCME, 
especially considering that it overestimates the cohesive energy of 
the ice crystal
by 58~meV/H$_2$O compared to 
experiment~\cite{santra2013accuracy}.
In order to
predict the 
abundance
of interstitial defects on basal plane ice surfaces,
it will be important to obtain more accurate theoretical values for these energy differences.

All methods applied here agree 
in
that the surface roughening process lowers the interaction energy, with close agreement between TIP4P/2005f, -200~meV, and SCME, -180~meV, while DFT/optPBE-vdW predicts weaker 
stabilization, -60~meV. Since the final state after the roughening process is characterized by a disordered, non-crystalline structure it appears likely that the weaker stabilization of DFT/optPBE-vdW 
derives from a different balance between electrostatic and dispersion interactions. 
The same holds true for the vacancy formation process on the Fletcher surface, where DFT/optPBE-vdW predicts the largest destabilization of the final state, which 
is characterized by a local disruption of the crystalline structure. 

\section{Conclusions}
We have 
carried out
AKMC simulations,
which are free from preconceived notions of possible transition mechanisms,
to shed light on possible 
molecular reordering processes of ice surfaces in 
the
cooler regions of Earth's atmosphere.
Structural and dynamical details of such processes are of fundamental importance to a range of 
atmospheric chemical reactions, but previous experimental and theoretical 
studies of ice surfaces have left many questions unanswered.
The results presented here demonstrate how the AKMC method can be used to investigate 
long timescale 
evolution
of complex systems such as the basal plane ice surface, 
without the need to apply biasing potentials or 
to increase temperature.
In particular, through automated saddle point searches,
AKMC can 
reveal
new and unexpected 
reordering mechanisms. 

Our observation of a series of
transitions that lead to a large distortion of the hexagonal lattice of O-atoms at
the proton-disordered surface is difficult to reconcile with existing 
helium atom scattering 
experimental data~\cite{glebov2000helium} 
which
indicate
a smooth basal plane ice surface at 90~K. 
The initial proton-disordered surface, constructed 
with
a commonly used method 
for generating proton-disordered ice 
simulation
cells, was metastable with respect to 
a 
disordering reconstruction which is hardly reversible due to 
the associated
increase 
in structural entropy. 
Further, the flipping process that occurred on the proton-disordered surface 
provides a possible low-barrier exothermic pathway to increase proton order.
Hence it appears unlikely that completely proton-disordered surfaces are representative 
models of real ice surfaces at low temperature.
Although by no means conclusive, the lack of any processes leading to
significant distortions of the lattice
on the Fletcher surface, despite 
$\sim$500 times longer simulation time
than for the proton-disordered surface, suggests that proton-order 
inhibits such processes at low temperature. 
Furthermore, our analysis of the event tables constructed by AKMC 
showed that the Fletcher surface 
is significantly more robust against local reordering 
transitions than the disordered substrate. In combination with the above mentioned experiments,
our simulation results, therefore, suggest that the surface is to a large extent 
stabilized by stripes of dangling 
H-atoms 
%H-bonds
as suggested by Fletcher~\cite{Fletcher:1992wi}.
This is particularly interesting when considering the 
relatively small difference, less than 9\%, 
in the energy 
of
the two surfaces.
Further work, however, will be required to reveal in more detail what type of patterns 
of dangling H-atoms might emerge and how 
they depend on temperature.

Our results also highlight the importance of carefully considering the surface proton 
order in simulation studies of ice, \textit{e.g.}, when investigating 
atmospheric chemical reactions. Using proton-disordered surfaces with a large amount 
of repulsive interactions between dangling 
H-atoms 
%H-bonds
may lead to 
unrealistic predictions of the energetics and reaction pathways.

On the other hand, the formation of proton-disordered patches on
largely proton-ordered
ice surfaces might 
be considered as rare events that can occur locally.
Previous experimental measurements may have been
insensitive to the type of roughening of crystalline order that we observe on the proton-disordered surface,
but its possible importance for the chemical 
reactivity of ice surfaces motivates further experimental 
efforts to characterize in more detail the surface structure of ice at low temperature.

Simpler reordering processes are observed on the ordered Fletcher
surface over a time interval of 49~$\mu$s, where
several molecules frequently rotate around 
their molecular axis and shift into the surface to become interstitial defects. 
Some of the
surface molecules are particularly susceptible to these 
processes and spend a significant portion of the  
simulated time
in the defect geometry.
Moreover, our observation of a transient process on the Fletcher surface lasting about 10~ns
which leads to a possible precursor state to a surface vacancy defect,
suggests that surface vacancies 
may form on a $\mu$s timescale even at low temperature,
\textit{i.e.}, around 100~K.
Future experimental and theoretical work 
would be required to quantify
the propensity of 
defects and elucidating how they affect the chemical reactivity of ice surfaces.

%\label{sec:Conclusions}

\subsection{Acknowledgement}
The authors would like to thank Dr. J-C. Berthet for helpful discussions.  
This work was supported by the Icelandic Research Fund 
and the University of Iceland Research Fund.
K.T.W. is supported by the Icelandic Research Fund through Grant No. 120044042.
L.J.K and H.M.C are funded by the European Research Council (ERC-2010-StG, Grant Agreement no. 259510-KISMOL)
and H.M.C. thanks The Netherlands Organization for Scientific Research (NWO) (VIDI 700.10.427).
A.P. thanks the COST Action CM0805 for enabling a research visit.
%\end{acknowledgement}

%\newpage %Just because of unusual number of tables stacked at end
\bibliography{hex_ice}% Produces the bibliography via BibTeX.

\end{document}